\documentclass[apl,twocolumn,showpacs,preprintnumbers,amsmath,amssymb]{revtex4}

\usepackage{graphicx}
\usepackage{dcolumn}
\usepackage{bm}
\usepackage[latin1]{inputenc}          
\usepackage[english]{babel}

\begin{document}

\title{Coherent acoustic phonons emission in copper driven by super-diffusive hot electrons} 

\author{M. Lejman}
\affiliation{Institut des Mol\'ecules et Mat\'eriaux du Mans, UMR CNRS 6283, 
Universit\'e du Maine, 72085 Le Mans,  
France.} 

\author{V. Shalagatsky}
\affiliation{Institut des Mol\'ecules et Mat\'eriaux du Mans, UMR CNRS 6283, 
Universit\'e du Maine, 72085 Le Mans,  
France.} 

\author{A. Kovalenko}
\affiliation{Institut des Mol\'ecules et Mat\'eriaux du Mans, UMR CNRS 6283, 
Universit\'e du Maine, 72085 Le Mans,  
France.} 


\author{T. Pezeril }
\affiliation{Institut des Mol\'ecules et Mat\'eriaux du Mans, UMR CNRS 6283, 
Universit\'e du Maine, 72085 Le Mans,  
France.} 

\author{V. V. Temnov}
\affiliation{Institut des Mol\'ecules et Mat\'eriaux du Mans, UMR CNRS 6283, 
Universit\'e du Maine, 72085 Le Mans,  
France.} 

\author{P. Ruello \footnote{Electronic address: pascal.ruello@univ-lemans.fr}}
\affiliation{Institut des Mol\'ecules et Mat\'eriaux du Mans, UMR CNRS 6283, 
Universit\'e du Maine, 72085 Le Mans,  
France.}

\begin{abstract}

Ultrafast laser excited hot electrons can transport energy supersonically far from the region where they are initially produced. We show that this ultrafast energy transport is responsible of the emission of coherent acoustic phonons deeply beneath the free surface of a copper metal sample. In particular we demonstrate that enough energy carried by these hot electrons over a distance as large as 220nm at room temperature in copper can be converted into coherent acoustic phonons. In order to demonstrate this effect, several configurations of time-resolved optical experiments of time of flight of coherent acoustic phonons and of hot electrons have been performed. 

\end{abstract}

\pacs{72.15.Lh, 63.20.kd, 73.50.Gr, 78.20.hc}
\maketitle

\section{Introduction}

Electrons moving at the Fermi surface at large velocities of about
$10^6$m/s determine the electrical conductivity and the heat
transport of metals. Since the electronic subsystem possesses a
small specific heat coefficient $C_{e}$ (as compared to that of
phonons $C_{i}$) and large intrinsic thermal conductivity $\kappa$
\cite{ash}, the electronic thermal diffusivity $D_{e}=\kappa
/C_{e}$ is also much larger than for the phonons. When the
electron-phonon coupling is weak, the Fermi electrons have the
ability to transport energy very rapidly and over large distances
as in the case of noble metals. These electrons possess a
large mean free path reaching several tens of nanometers
\cite{crow}. The dynamics of Fermi electrons in metals have been
investigated in the time domain during the last twenty years consecutively to
the development of femtosecond (fs) laser technology. By exciting
non-equilibrium electrons with an intense ultrashort pump pulse
and monitoring the subsequent evolution of time-dependent optical
properties of the sample by a time-delayed ultrashort probe
pulse, the electrons dynamics have been well characterized
\cite{bror1,bror2,ees,elsa,schoen,delfatti1,delfatti2,juhas}. The
wavelength dependence of time-resolved optical  reflectivity
signals across the d-band of metals allowed to directly probe the
'smearing' of the Fermi surface caused by transient electron
heating. To model the electron dynamics the two-temperature model
\cite{schoen,ees,juhas} or non-thermal model \cite{gus1,tas} were
proposed. The difficulty of determining experimentally the
dynamics of electrons relaxation in pump-probe experiment lies on
the fact that hot electrons not only loose their energy through the
interaction with the phonons but also because they escape from the optical
skin depth region of excitation thus making the analysis much more
delicate \cite{bror1,hohl}. Therefore, when both pump and probe
pulses are incident on the same surface (front-front
configuration), the decaying transient optical signal is not only
affected by electron-phonon relaxation but also due to the escape
of hot electrons out of the skin depth region. Optical experiments with
probe pulses incident from the back side of laser-excited films
permitted to observe the ultrafast transport of hot
carriers from the front to the back surface on a sub-picosecond
time scale \cite{bror1,juhas}. These studies have demonstrated
that over-heated electrons can ballistically transport energy at a
very short time scale \cite{bror1}. In gold, ballistic transport of electrons up to $300$ nm has been
reported \cite{juhas}. This ultrafast transport of energy exceeds
by more than one order of magnitude the typical optical skin depth
in a metal. These effects have been extensively studied for gold
and silver \cite{juhas,bror1}.

\begin{figure}[!b]
\centerline{\includegraphics[width=9cm]{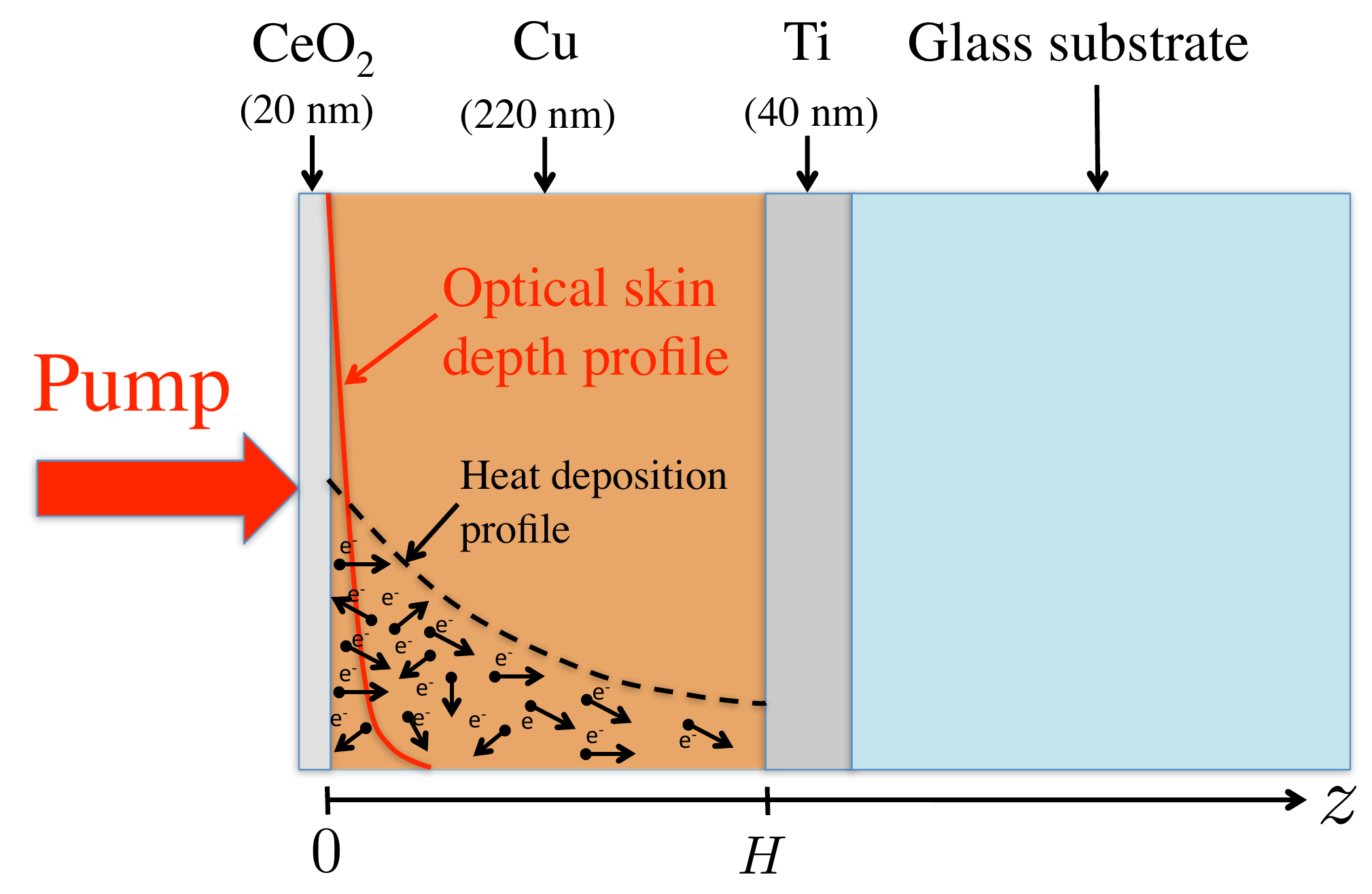}}
\caption{\label{fig1} (color online)  The femtosecond laser radiation is absorbed over the skin depth region (absorption profile depicted by a solid red line) while superdiffusive hot electrons transport energy up to the copper/titanium interface (profile depicted in dashed line). Through the electron-phonon coupling process (thermoelastic process in that case), coherent acoustic phonons fronts are generated not only at the copper surface ($z \sim 0$) but also at the copper/titanium interface ($z \sim H$) }.

\end{figure}

These super-diffusive hot electrons relax by
electron-phonon interaction and consequently give rise to acoustic phonons
emission in the picosecond time scale through the thermoelastic process (ultrafast lattice heating). The typical spatial
extension L  of the emitted
acoustic phonon wavepacket is of the order of the distance over which the
electrons deposit their energy, which is large when
electron-phonon coupling is low. The duration of the picosecond
acoustic pulse is then simply $\tau \sim L/V_{S}$ where $V_S$ is the sound velocity. This effect has clearly been illustrated in picosecond acoustic
experiments, where the spatial extension L of the thermoelastically induced acoustic phonons wavepacket
 in gold was found to exceed by an order of
magnitude the optical skin depth. This gave rises to spatial and
temporal broadening of the emitted acoustic pulse and subsequent
reduction of the emitted acoustic phonons frequencies. Several
experimental signatures of  diffusive electrons
transport can be inferred from the broadening of the measured picosecond
acoustic pulse \cite{wri0,wri1,sai,pez1,pez2}. On the contrary, metals having large electron-phonon coupling like
Cr, Ni \cite{sai} Co
\cite{Temnov12NPhoton6,TemnovAcoustoPlasmonics}, Ti \cite{lin},
the diffusion length of non-equilibrium electrons becomes much shorter,
typically of the same order of magnitude as the optical skin depth
itself. As a consequence, the picosecond acoustic pulse duration
becomes much shorter that those observed for noble metals.\\
Even if this ultrafast supersonic transport of energy by electrons has been
taken into account to describe the spectrum of photogenerated
acoustic phonons in metals \cite{wri0,wri1,sai,gus2} only one
experimental study of time of flight of coherent acoustic phonons has permitted to evidence more directly the ultrafast transport of hot electrons. Tas and Maris \cite{tas},
have discovered the photogeneration of coherent acoustic phonons at the two opposite sides of a $70$ nm thick aluminum film induced by ultrfast transport of hot carriers. Following the first observation made by Tas and Maris here we
report direct evidence of that this effect is very efficient in copper and we obtain clear signature of the contribution of ultrafast transport of hot carriers and its consequence on coherent acoustic phonons emission. With an
extended set of experimental configurations we show that supersonic
heating in copper can transport enough energy to generate coherent
acoustic phonon deeply beneath the metal free surface. In particular, we
show that this process is efficient over quite large
distances as large as $\sim 220$ nm at room temperature.
We show that femtosecond pump beam interacting with a copper film excites not only coherent acoustic phonons at the front surface ($z\sim0$ in Fig. \ref{fig1}), but also, thanks to ultrafast transport of energy by hot electrons, coherent acoustic phonons up to the copper/titanium interface ($z\sim H$). Several experiments where we demonstrate the efficiency of this phenomenon by probing the emitted coherent acoustic phonons at the front and at the back surface have been performed in this study. \\
Understanding the hot electrons energy transport in several non-equilibrium photo-induced processes which aim at controlling or/and modifying the charge or spin order in materials for example  \cite{PT0,PT1,PT2} still remains an important issue  Moreover, recent reports even propose to inject hot electrons in materials to induce phase transitions \cite{richard}.

\section {Experimental methods and samples}

\begin{figure}[t!]
\centerline{\includegraphics[width=9cm]{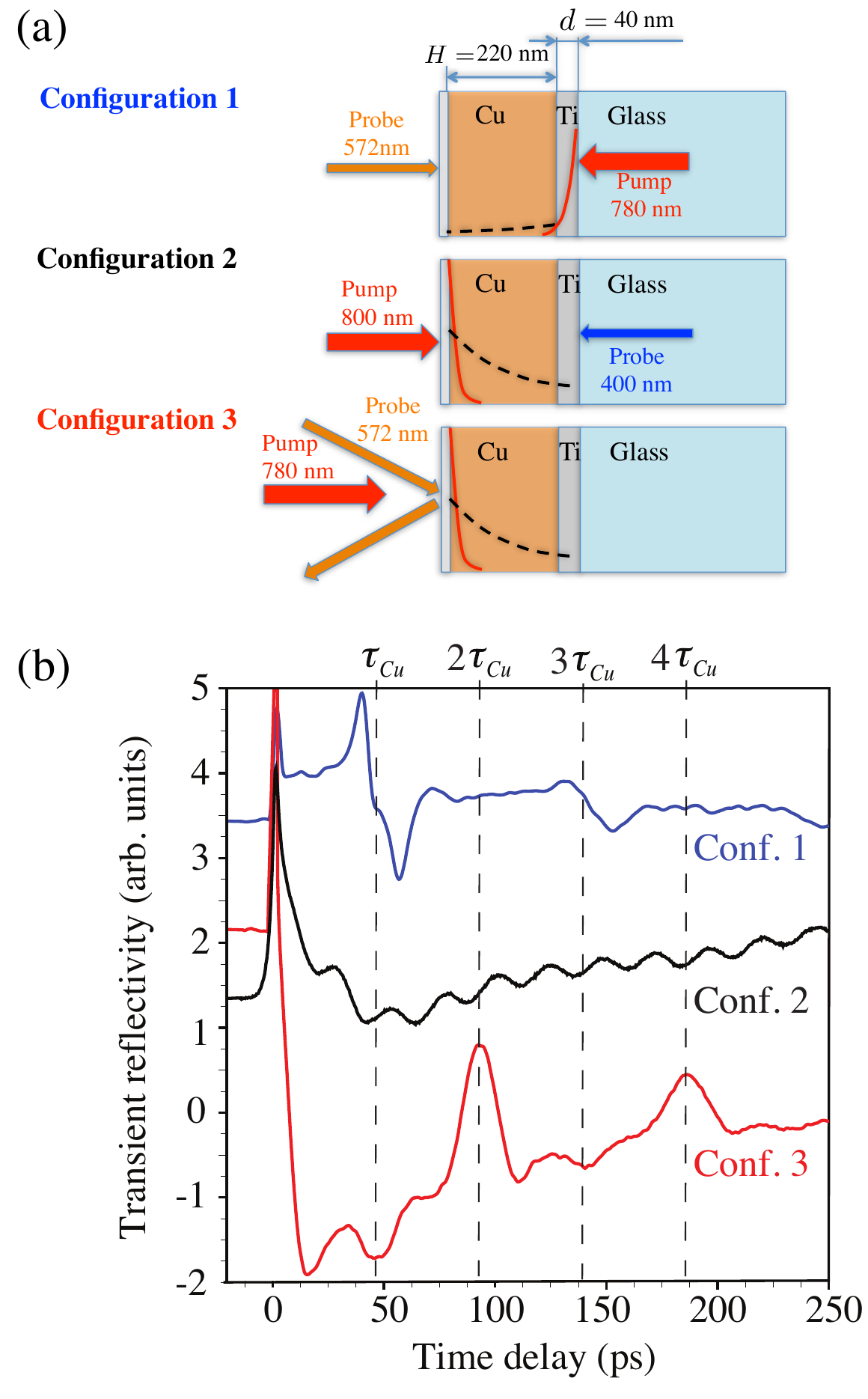}}
\caption{\label{fig2} (color online)  (a) Description of the three different front-back and front-front experimental configurations. The optical excitation is depicted as red lines profiles indicating the evanescent light penetration in the metals (copper or titanium). On the other hand, the dashed lines illustrate qualitatively the profile of energy deposition of hot electrons consecutively to their super-diffusive transport. (b) Time resolved optical reflectivity measurements according to the three configurations (1, 2 and 3). The time constant $\tau_{Cu}=H/V_{Cu}$ corresponds to the time of flight of the acoustic wavepacket through the copper layer of thickness $H$ at the acoustic speed $V_{Cu}$}.

\end{figure}

Depending on the phenomenon we study, the pump and probe experiments were performed according to two femtosecond setups. A first setup employs a mode-locked Ti:Sapphire cavity dumped femtosecond laser of 500kHz laser pulse repetition rate. The laser beam output at 800nm is separated into a pump and a probe beam by a polarizing beam splitter. The probe beam passes through a second harmonic generation (SHG) BBO crystal in order to obtain a probe wavelength of 400nm. The pump is chopped by an acousto-optic modulator at around 100kHz. Both beam are focus by $\times 10$ microscope objectives. The transient optical signal recorded with a balanced photodiode is processed with a lock-in-amplifier at the pump modulation frequency. The second setup employs a Ti:Sapphire laser of tunable wavelength. The laser beam output is splitted into a pump beam and a probe beam that pumps an Optical Parametric Oscillator (OPO) having an intra-cavity SHG doubling crystal. The OPO allows the probe wavelength to be tunable from 520nm to 600nm with a step of around 1nm. The pump beam is chopped with an electro-optic modulator at around $100$ kHz and the transient optical signal is recorded with a balanced photodiode and processed with a lock-in-amplifier at the pump modulation frequency.\

We have chosen to study the supersonic transport of electronic energy in copper for different reasons. First, electrons in copper are known to have large mean free paths of the order of a hundred of nanometers. Copper has a small electron-phonon coupling constant with $g= 50\times10^{15} W.m^{-3}.K^{-1}$ \cite{gcu1,bror2,elsa}. Secondly, copper has a d-band optical transition centered at 574nm which lies within the optical range of our OPO (520-600nm). Consequently, we have proper OPO output probe wavelengths to better investigate the dynamic of hot electrons around the Fermi energy. A copper film has been deposited onto a 40nm thick titanium film lying onto a glass substrate of 1 mm thickness (see Fig. \ref{fig1}). In comparison with copper, titanium exhibits a much larger electron-phonon coupling constant ($g \sim 200-500.10^{15}$ $W.m^{-3}. K^{-1}$) \cite{bror2} and permits to generate quite high frequency coherent acoustic phonons \cite{lin} and can be therefore consider as a so-called "good" thermoelastic transducer. In order to protect the copper free surface from oxidation, a $20$ nm thin layer of cerium oxide was deposited at last.

\section {Results and discussions} \label{results}

\subsection {Generation of coherent acoustic phonon by supersonic hot electrons}

Following the classical picosecond acoustic scheme, we performed first experiments in front-back configuration 1, see Fig. \ref{fig2}(a), in this case the laser pump excites the titanium film and through thermoelastic process leads to the generation of a picosecond acoustic pulse that propagates within the copper film and is detected at the front surface of the copper film. The transient optical reflectivity signal exhibits different features. We can easily observe some periodic events at time around $t \approx 45$ps, $145$ps and $230$ps. Two of these successive events are separated in time by the value very close to $2\tau_{Cu} \sim 2H/V_{Cu}$. This time exactly corresponds to a round trip time of the longitudinal acoustic phonons in copper. Consequently, we can attribute these events to the acoustic pulse photogenerated in the titanium film and travelling back and forth within the copper film at the longitudinal acoustic speed $V_{Cu}=4730$m/s \cite{HB}. Since the acoustic impedance $Z=\rho \times V_{S}$ of titanium is smaller than that of copper, see Table \ref{tab1}, then the successive acoustic pulses display the same transient reflectivity sign.

The generation of short longitudinal acoustic pulses in metals and detection of successive acoustic echoes have been already extensively described in picosecond acoustics and is not of concern here \cite{tom1,devos1,perrinnatur,wri0}. However, we can note that a sharp offset of the transient reflectivity signal appears before the first acoustic pulse detection. This sharp peak, of about 3 ps of duration, appears prior to the arrival of the acoustic phonons coming from the titanium film, within a delay of $\tau_{Cu}=H/V_{Cu}$. As a consequence, this sharp peak corresponds to the so-called zero time which defines the onset of pump excitation. Such a peak has already been observed in similar time of flight experiments in gold \cite{bror1} and silver \cite{juhas} and is attributed to the arrival of super-diffusive laser-excited electrons. We will discuss the nature of these hot electrons in section \ref{secbal}.  

\begin{table*}[t!]
\begin{tabular}{ccc}
\hline
Some Physical Properties of copper and titanium  & &\\
\hline
\hline
Optical properties \cite{optcopper1,optcopper2}& &\\
\hline
 Wavelength(nm)&        skin depth (nm) & \\
 &  Copper  & Titanium   \\

 400nm & $\sim 14.5$             &$\sim 14.4$                              \\
  572nm & $\sim 17.5$             &$\sim 16.8$                             \\
   780nm & $\sim 12.4$             &$\sim18.8$                            \\
    800nm & $\sim 12$            &$\sim19.3$                           \\
  
\hline
\hline
Mechanical properties & &\\
\hline
                 Sound velocity $V_{S}$ ($m/s$) & $4730$            &$5090$                           \\   
                       Density $\rho$ ($kg/m^3$) & $8960$            &$4500$                           \\       
 \hline
\hline
Transport properties  & &\\       
\hline
Electron-phonon coupling constant \cite{gcu1,bror2,elsa} $g$ ($10^{15} W.m^{-3}.K^{-1}$) & 20-60 & 200-500\\
Heat conductivity \cite{HBCe} $\kappa$ ($W.m^{-1}.K^{-1}$) & 400 & 80         \\                             

\hline

\end{tabular}
\caption{\label{tab1}  Some physical properties of copper and titanium}
\end{table*}

In the second front-back configuration the pump beam excites the copper film and the probe beams detects the photo-generated acoustic pulse in the titanium film through the transparent glass substrate (configuration 2 in Fig. \ref{fig2}(a)). In that case we observe a transient reflectivity signal composed of a fast rising peak followed by an oscillating component (Fig. \ref{fig2}(b)). The frequency of this oscillating component corresponds to the Brillouin frequency in glass. As in any Brillouin scattering measurement, the Brillouin period is given by $T_{B}=\lambda / (2V_{glass}\sqrt{n^2-sin^2(\theta)})$, where $\lambda=400nm$, $V_{glass}=5500$m/s, $n\approx1.47$ and $\theta=0$ are the probe wavelength, the longitudinal sound velocity in glass, the refractive index of glass at the probe wavelength $\lambda$ and the incidence probe beam angle. In our case we obtain $T_{B}= 24$ ps which is in close agreement with the experiment. Brillouin scattering in glass is well known and has been observed many times in several picosecond ultrasonics experiments \cite{brill}. The most important result here concerns the fact that the Brillouin oscillations starts before the arrival time of the acoustic pulse photogenerated at the free surface of copper (see Fig. \ref{fig2}(b)). We can also observe a kink in the signal at a time close to $\tau_{Cu}=H/V_{Cu}$ which corresponds the traveling time of an acoustic wave photogenerated at the copper surface to reach the substrate. These results demonstrate that, without directly optically exciting the titanium film, it is possible to  generate nearly immediately (order of the picosecond time scale) acoustic phonons in the titanium film and the glass substrate by a non-optical process. We claim that this process is the due to a mechanical stress triggered by the energy transported by super-diffusive hot electrons coming from the free surface of irradiated copper film. 
These observations are strongly supported by the last set of experiments, see front-front configuration 3 in Fig. \ref{fig2}(b). In that case both pump and probe are focused at the front surface of the copper film, which means that acoustic phonons are directly laser excited at the front copper surface and, after an acoustic round trip through the copper film, they are detected at the front surface. These acoustic echoes are well distinguished at time $2\tau_{Cu} \approx 90$ps$=2H/V_{Cu}$ and $4\tau_{Cu} \approx 180$ps$=4H/V_{Cu}$ as expected. The sign of these successive acoustic pulse remains the same since that acoustic pulse undergoes two acoustic reflections onto the free surface and on the titanium substrate which have lower acoustic impedance than that of copper. The time width of this acoustic pulse is around $30$ps which is much larger than the characteristic time associated to the traveling time acoustic phonons across the optical skin depth ($\delta_{skin}$) of the laser pump photon in copper, that is  $\tau_{skin} \sim 2\delta_{skin}/V_{Cu} = 12$nm$/4730$m/s $\sim 3$ps. This broadening is similar to what has been observed by Wright et al \cite{wri1} and is due to thermal and electronic diffusion in copper. We will discuss this electrons transport in part \ref{num}. More importantly, in addition to these acoustic pulses and acoustic echoes, we clearly observe on the transient reflectivity signal a periodic pulse of the optical reflectivity occurring first at time $\tau_{Cu} \approx H/V_{Cu}$ and appearing every $2\tau_{Cu}=2H/V_{Cu}$. The fact that these events occur with a time period  corresponding to the time of flight of longitudinal acoustic phonons for a round trip within the copper film, shows that these detected signals are also of acoustic nature. However, since the first signal appears at a time $\tau_{Cu}\approx H/V_{Cu}$, this indicates that the acoustic phonons pulse front has been generated at a distance of $H$ beneath the copper free surface. Finally, since the substrate cannot be reached by the  pump beam radiation, it means that acoustic phonons are then produced by an non optical ultrafast process, i.e. a super-diffusive transport of energy up to the copper/titanium interface mediated by hot electrons. Consistently with the results shown for configuration 2 (detection of Brillouin signal in the glass substrate), these additional results confirm the ultrafast excitation of coherent acoustic phonons  far away from the optical skin depth of pump radiation in copper. 
An important feature obtained in configuration 3 concerns the opposite sign between the detected acoustic pulses originating from direct laser excitation (S) or from superdiffusive hot electrons (B) in the transient reflectivity signal, see Fig. \ref{fig3}(b). A straightforward analysis of the acoustic reflection coefficients indicates that during the excitation of both acoustic pulses (S) or (B), the strain field has the same sign. Since we know that (S) acoustic pulses are driven by thermoelasticity (negative photoinduced stress \cite{wri1}), it is an indication that the same process takes place at the copper/titanium interface during the electron-acoustic energy conversion which generates the (B) acoustic pulses.
It is worth to be noted that picosecond acoustic pulses laser excited in a 235 nm copper film were already investigated by Wright et al \cite{wri1} in the past.  However, probably because of poor signal to noise ratio, they did not observe the (B) acoustic pulse excited by superdiffusive hot electrons. Since it is known that grain boundaries can deeply affect the transport of hot electrons \cite{juhas},  another explanation for not measuring the (B) acoustic pulse could be that the quality of the copper film was not good enough.

We will see in the following section that we can numerically reproduce our observations considering an instantaneous distribution of the energy in the whole copper film driven by superdiffusive hot electrons. 

\subsection {Numerical simulations}\label{num}

\begin{figure}[t!]
\centerline{\includegraphics[width=8cm]{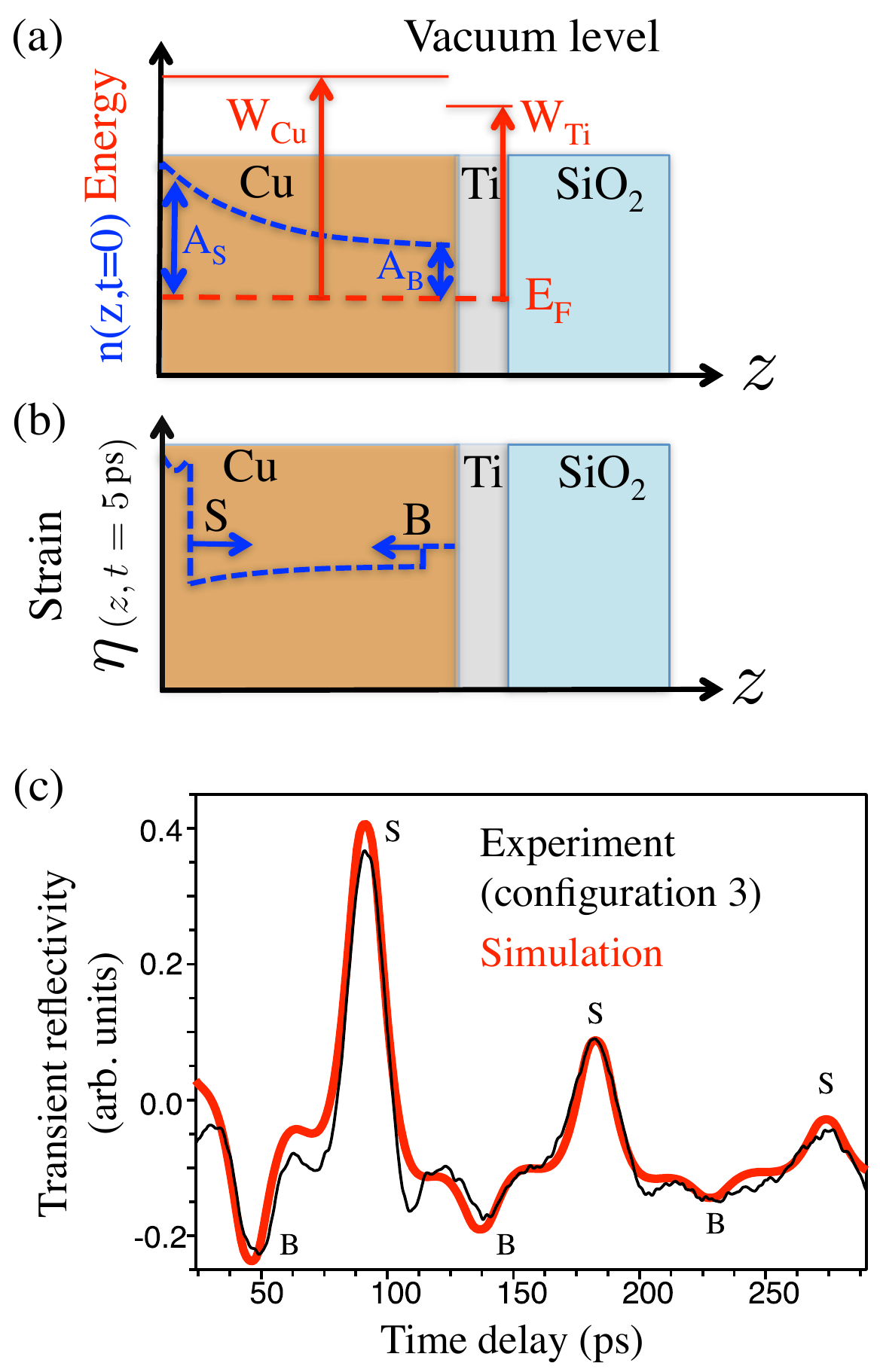}}
\caption{\label{fig3} (color online) (a) Sketch of the hot electron distribution (blue dashed-lines) after their ultrafast diffusion. The work function of electrons (energy scale in red color) in copper being larger than that in titanium, electrons remain in the copper film. The vacuum level for each metal is depicted as a solid horizontal red line. (b) Sketch of the emitted acoustic pulse (arb. units) with the acoustic fronts coming from the surface denoted as S and that from the back of the copper film denoted as B. (c)  Comparison between experimental signal obtained in the configuration 3 (see Fig. \ref{fig2}(a)) and our simulation}. 
\end{figure}

In order to gain understanding on the superdiffusive hot electron phenomenon and its relation to the emission of coherent acoustic phonons, we have performed some numerical simulations that quantitatively reproduce our experimental datas. The description of the hot electrons in depth energy distribution is still a matter of debate and has been already modeled either according to an athermal model \cite{tas} or according to an electron diffusion approximation \cite{wri1}. Ballistic or diffusive signature of hot electrons transport is a current matter of discussion \cite{schneider}.  In our analysis, we have first simply considered that the thermoelastic stress $\sigma =-3B\beta \Delta T$ \cite{tom1}, where B, $ \beta$ and $\Delta$ T are the bulk modulus, the thermal expansion coefficient and the lattice temperature increase due to the energy transfer from the electronic to the phonon subsystem, applied to the lattice is instantaneous compared to the characteristic time of the detected coherent acoustic phonons. We remind that at the phonons frequencies we are dealing with ($<300GHz$), the deformation potential for metals has a little contribution in this frequency range \cite{wri0,wri1,sai,gus4} and can be neglected. Thus the driving mechanism of coherent acoustic phonons emission is mainly the thermoelastic process. In our simulation we have adopted a phenomenological approach based on the hot electron diffusion approximation established earlier \cite{wri1}. It has been shown that the strain profile could be depicted following an exponential shape with $\eta(z,t) \propto  exp(- z/\xi)$ \cite{wri1}, where $\xi$ is the characteristic length of energy transport by hot electrons with $\xi=\sqrt{\kappa / g}$ and $\kappa$ and $g$ are the heat conductivity and the electron phonon coupling constant, see Table \ref{tab1}. We have considered that this hot electrons distribution is instantaneous in our model (Fig. \ref{fig3}(a)). The emitted acoustic strain $\eta(z,t)$ is then depicted in Fig. \ref{fig3}(b) for a time $t\sim$ 5 ps where we take into account the acoustic reflection at the free surface and at the titanium layer. Considering that the probability of electron transfer from copper to titanium is limited due to the difference of the work function with $W(Cu) \sim 4.5-5.1eV$ and $W(Ti) \sim 3.8-4eV$, we have neglected the electronic energy deposition into the titanium film.

The numerical simulation of the transient reflectivity shown in Fig. \ref{fig3}(c) has been carried out following the established photoelastic model of the detection process \cite{tom1,perrinnatur,gus3,wrightJAP} where the measured transient reflectivity is $\Delta R/2R =Re(\Delta r/r)$, with : 

\begin{eqnarray}
\label{detect}
\ \Delta r/r_{}&=&2ik_{0}u(0)\nonumber\\
&+& ik_{0}p_{e}\frac{4n_{probe}}{1-{n_{probe}}^2}    \int_{0}^{H}{\eta (z,t)e^{2ik_{probe}zdz}}\nonumber
\end{eqnarray}

$k_{0}$, $u(0)$, $n_{probe}$ and $p_{e}=dn_{probe}/d\eta$ being the probe light wave vector in vacuum, the free surface displacement of copper,  the refractive index of copper at the probe wavelength respectively and the photoelastic coefficient. The optical refractive index at the probe wavelength $572nm$ is $n_{probe}=0.7+3.5i$ \cite{optcopper1,optcopper2}.  To reproduce the broadening of the acoustic pulse coming from different physical mechanism (electron diffusion, acoustic phonon scatterings and transit time of the acoustic pulses in the thin cerium oxide coating layer) we have numerically broadened the acoustic front by applying a cut-off frequency ($300$ GHz). In particular, broadening of acoustic pulse by scattering at grain boundaries adds some delay in the time propagation of acoustic phonons \cite{maz}. \

The experimental results (Fig. \ref{fig3}(c)) indicate that the ratio between amplitudes of so-called the S-acoustic pulse and B-pulse is around 3. As shown in the following with a simple model, we can extract from this ratio a direct evidence of the ultrafast transport of hot electrons in copper and its determining role in the coherent acoustic phonons emission process. First of all, it is to be mentioned that in picosecond acoustics, when the detection of coherent acoustic phonons takes place in an opaque system like a metal, only the high frequency components of the acoustic pulse are revealed as it has been theoretically and experimentally established earlier \cite{babi1}. This means that we mainly probe the magnitude of the front of the incident acoustic pulse. Furthermore, in the thermoelastic model, the amplitude of the acoustic pulse is proportional to the lattice temperature increase which is proportional to the local hot carriers density. Therefore, we can assume that the acoustic strain amplitude follows the hot electron density. Considering, the initial profile of hot carriers with $A_{B}/A_{S}=exp(-220 nm/ \xi)$ (see Fig. \ref{fig3}(a)) where $\xi$ is the characteristic diffusion length of the hot electrons, and taking into account the rules of acoustic reflection at boundaries, we can then write with a reasonable approximation, that the amplitude ratio of (S) and (B) detected acoustic pulses follows  ${2A_{S}\mid R_{Cu/Ti}\mid}/{A_{B}(1+ \mid R_{Cu/Ti}\mid)}/ \sim 3$. $ R_{Cu/Ti}$ is the acoustic reflection at the copper/titanium interface. From this equality, we obtain $A_B / A_S=exp(-220 nm/ \xi)  \sim 1/6$. Then, we estimate the hot electrons diffusion length to $\xi \sim 120$ nm. Finally, with the tabulated value of heat conduction in copper, this leads to an electron-phonon coupling constant $g \sim \kappa / \xi^2 \sim  28$ $ 10^{15} W. m^{-3}.K^{-1}$ in agreement with the literature \cite{bror2}. As a summary, both the numerical description and the analytical model show that the detected (B) acoustic signal comes from the ultrafast transport of hot carriers.

Even if the general shape of the signal is pretty well reproduced by our numerical simulation, it clearly appears that there is a slight time delay of 1-2 ps between the arrival time of the experimental B pulse compared to the calculated one. This effect could be due to the finite time propagation of the hot electrons across the $220$ nm thick copper film. In our phenomenological numerical model, we have considered that the photo-induced stress is immediately applied to the entire copper film. In reality, the characteristic time of hot electron diffusion in copper is given by $\tau \sim l^2/4D$ \cite{wri1}, where $l$ and $D$ are the distance of diffusion and the diffusion coefficient with $D=\kappa / C_{e}$ where $C_{e}=96$ $T_{e}$\cite{HBCe} is the heat capacity of electrons in copper and $T_{e}$ the electronic temperature. In our experiments, we used moderate pump fluence such that the hot electron temperature is estimated at $T_{e}\sim 400$ $K$. With the tabulated heat conductivity value of $\kappa$, we get $D \sim 10^{-2}$ $m^{2}.s^{-1}$. Consequently, the characteristic diffusion time of hot electrons is $\tau \sim 1.2$ $ps$, which is in agreement with the delay accumulated by the B-pulse compared to the ideal case of immediate application of the photo-induced thermoelastic stress in copper.

\subsection {Probing ultrafast transport of hot electrons}\label{secbal}

\begin{figure}[t!]
\centerline{\includegraphics[width=8cm]{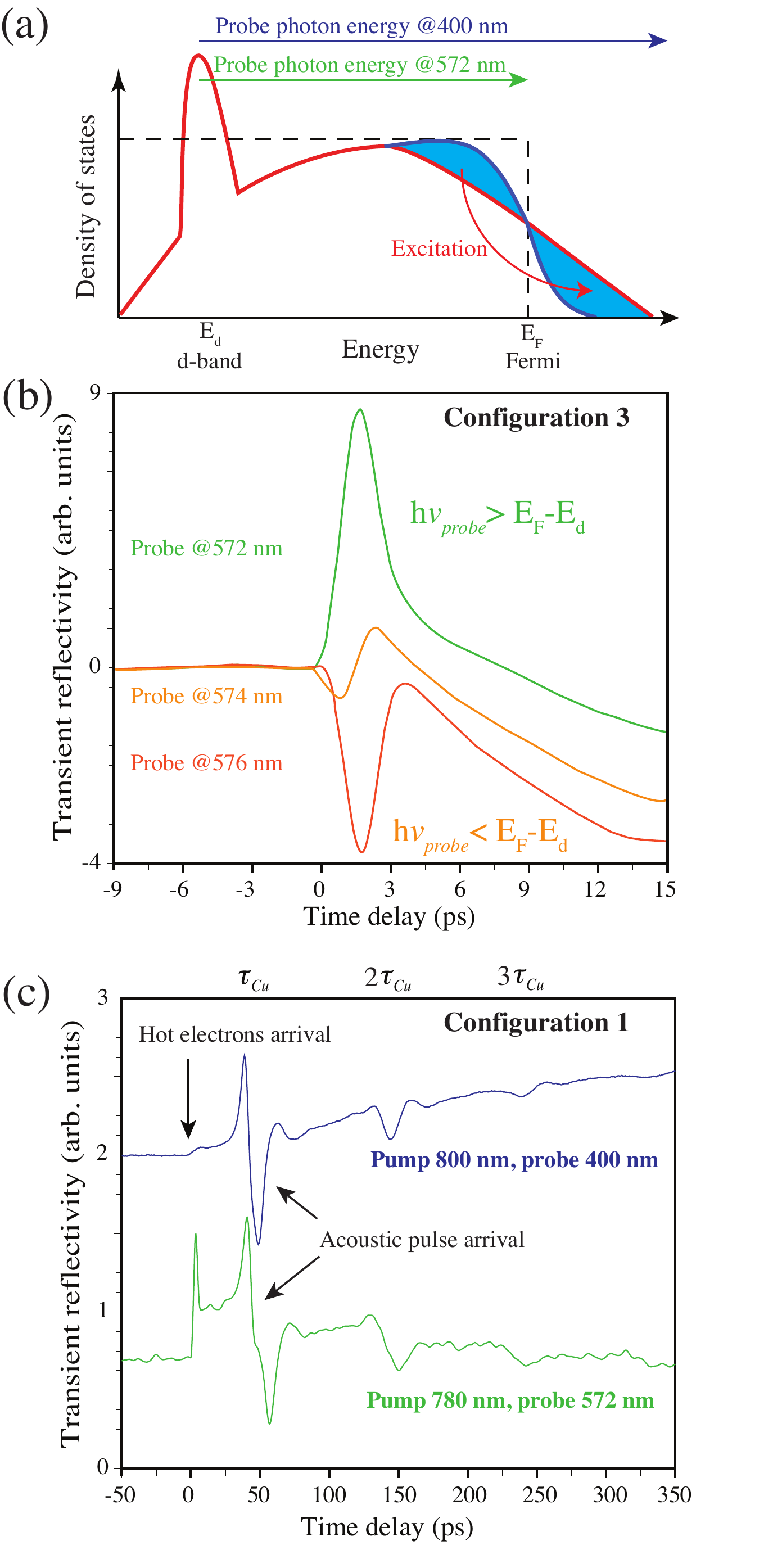}}
\caption{\label{fig4} (color online)  (a) Sketch of electronic density of states for copper. When electrons are excited, it gives rise to a Fermi distribution smearing as indicated. In a given place of a metallic crystal, this electrons excitation  can be reached either by direct optical excitation or by the transport of hot carriers towards this place. The Fermi distribution smearing can then be probed with different probe wavelengths. (b) Effect of the probe wavelength on the time resolved optical reflectivity obtained according to the configuration 3. Probing the hot electron distribution with a probe photon larger or smaller than $E_F-E_d$ leads to an increase or decrease of the electronic component of the transient optical reflectivity.(c) Probe wavelength dependence of the transient optical reflectivity obtained in configuration 1. The electronic response at $t\sim 0$ ps exhibits a huge wavelength dependence in accordance with theory. }
\end{figure}

In the previous section we have observed that picosecond acoustic pulse can be generated by super-diffusive energy transport involving hot electrons. In the last section, we discuss the optical signature of this ultrafast hot electrons transport. In the experiments performed according to configuration 1 and 2, a disturbance of the optical reflectivity signal at $t\sim 0$ps was systematically observed prior to the arrival of the acoustic pulses at $\tau_{Cu}=H/V_{Cu}$. This signal at $t=0ps$ comes from the ultrafast hot electrons that can travel a distance of more than 200nm within around $1.2$ ps as estimated in the previous section \ref{num}. This ultrafast transport of electrons has already been observed in gold for example \cite{bror1}. For the pump and probe experiment conducted according to the configuration 1, even if the main optical energy is deposited inside the titanium film (see Table 1), around  $10\%$ of the light goes through the titanium film (this is depicted by the profile of light penetration indicated by a red line in Fig. \ref{fig2}(a)). Due to the high electron-phonon coupling in titanium metal, and even if the work function of titanium is smaller than that of copper (see Fig. \ref{fig3}(a)), we assume that a few of photo-excited titanium electrons actually are transferred into the copper film. As a consequence, slight direct photo-excitation of copper electrons occurs only. These electrons travel through the copper film very rapidly and are detected at the front copper surface. These hot electrons, because they are not totally thermalized with the lattice subsystem, have a temperature which is larger than that of phonons. This temperature difference induces a smearing of the Fermi sphere \cite{schoen,ees,elsa,bror1}. Consequently, the density of occupied states is modified in the vicinity of the Fermi level as depicted in Fig. \ref{fig4}(a). This effect has been extensively studied previously in the case of gold and silver in front-front configuration only \cite{ees,delfatti1,delfatti2,schoen, bror1}. By choosing a proper probe wavelength, this Fermi distribution smearing can be directly probed. It has been shown that when the probe photon energy is smaller than the energy between Fermi energy and the d-band level (ie. $E_{probe}<E_{F}-E_{d}$), the thermomodulation of the optical reflectivity is characterized by a decrease of the optical reflectivity ($\Delta R<0$), because more probe photons can be absorbed since more empty states appears below Fermi level. On the contrary, when the probe photon energy is larger ($E_{probe}>E_{F}-E_{d}$), there is an increase of the transient optical reflectivity ($\Delta R>0$). We have experimentally checked the occurrence of this optical effect in the case of copper and we indeed observed a change of the sign of the transient optical reflectivity in the first picoseconds time range ($t\sim 0$ ps) in the vicinity of the d-band transition. The experiments conducted according to the front-front configuration 3  clearly evidences the wavelength dependence of the electronic response as expected (see Fig. \ref{fig4}(b)). This observation is in full accordance with the results of Eesley \cite{ees}. 

Measurements conducted according to the front-back configuration 1, at two different probe wavelengths 572 nm and 400 nm, have been performed as well, see the corresponding results in Fig. \ref{fig4}(c). First of all, we can observe that the detected acoustic pulses at time $\tau_{Cu}$, 3$\tau_{Cu}$ display different shapes. This effect is well known \cite{devos1,babi1}, it is due to the modification of the photoelastic coefficients versus probe wavelength.. More interestingly, a strong wavelength dependence related to the detection of ultrafast hot electrons is evidenced at time $t\sim$ 0 ps.  While at 572 nm probe wavelength, we detect a strong and sharp peak, the detection of hot electrons arrival at 400 nm probe wavelength is very weak and only a slow jump appears on the transient reflectivity signal (at time $t\sim 0$ ps). The sharp rising peak observed at 572 nm probe wavelength is in agreement with the expected increase of the probe optical reflectivity that should appear when probing copper metal with a photon energy $E_{probe}> E_{F}-E_{d}$, see Fig. \ref{fig4}(a). At 572 nm probe wavelength, even if only a small amount of copper electrons reach the free surface, they are efficiently measured because the detection occurs at the vicinity of a d-band transition which is very sensitive to any subtle change of the electronic distribution. On the contrary, for probe photon energy much larger than $E_{F}-E_{d}$, for instance at 400 nm probe wavelength, the inter-band transition does not drive anymore the optical response of the metal. This is the reason why we observe a weak slow jump of the optical response, as already observed in thermo-reflectance modulation spectroscopy \cite{optcopper2}

As a summary, this investigation shows that the Fermi distribution is smeared within a picosecond time scale due to the arrival of hot carriers at the copper free surface. This provides an additional illustration that ultrafast hot electrons actually are able to travel the copper film very fast. It is worth to mention that in the configuration 1 when a small amount of copper electrons are excited, it is enough to detect them at the copper surface after they travelled the copper film because the detection occurs in the vicinity of a d-band transition which is very sensitive to any subtle change of the electronic distribution. However, we did not observe any conversion of their energy into mechanical energy to emit for example an acoustic pulse at the interface between the protective layer of cerium oxide and the copper film. We did not observe any acoustic signal at a time of around $2\tau_{Cu}=2H/V_{Cu}$ and $4\tau_{Cu}=4H/V_{Cu}$ in the configuration 1. This means that the electronic energy conversion of the few hot electrons reaching the free surface into mechanical energy does not efficiently drive any measurable acoustic pulse.

 \section {Conclusion}

In this study, by conducting several experiments of time of flight of picosecond acoustic phonons, we have clearly evidenced a discrepancy in the excitation process of coherent acoustic phonons between strong (Ti) and weak (Cu) electron-phonon coupling metals. In the latter one, we have shown that hot electrons can transport energy and trigger coherent acoustic phonon emission over a distance larger than 200nm at room temperature which is ten times larger than the optical skin depth in copper. Probe wavelength dependence of the detection process has permitted to clearly reveal that hot electrons can travel across a thick copper film within a very short time ($\sim  1ps$) for a copper film thickness of $220$ nm. Due to this efficient transport of energy by hot carriers, we could imagine to employ them to trigger different dynamics with a gun of hot electrons  like phase transitions, electric dipole or spin reorientation. Recent observations of this possible hot-electrons induced phase transition have been proposed \cite{richard} as well as demagnetization driven by an injection of super-diffusive hot-electrons in magnetic materials \cite{esch}.\

We thank G. Vaudel and V. Gusev for helpful discussions.

\newpage

\end{document}